\documentclass[12pt]{article}

\usepackage{amssymb}

\def\fnote#1#2{\begingroup\def\thefootnote{#1}\footnote{#2}\addtocounter{footnote}{-1}\endgroup}

\def\inbar{\vrule height1.5ex width.4pt depth0pt}
\def\IB{\relax{\rm I\kern-.18em B}}
\def\IC{\relax\,\hbox{$\inbar\kern-.3em{\rm C}$}}
\def\ID{\relax{\rm I\kern-.18em D}}
\def\IE{\relax{\rm I\kern-.18em E}}
\def\IF{\relax{\rm I\kern-.18em F}}
\def\IG{\relax\,\hbox{$\inbar\kern-.3em{\rm G}$}}
\def\IH{\relax{\rm I\kern-.18em H}}
\def\II{\relax{\rm I\kern-.18em I}}
\def\IK{\relax{\rm I\kern-.18em K}}
\def\IL{\relax{\rm I\kern-.18em L}}
\def\IM{\relax{\rm I\kern-.18em M}}
\def\IN{\relax{\rm I\kern-.18em N}}
\def\IO{\relax\,\hbox{$\inbar\kern-.3em{\rm O}$}}
\def\IP{\relax{\rm I\kern-.18em P}}
\def\IQ{\relax\,\hbox{$\inbar\kern-.3em{\rm Q}$}}
\def\IR{\relax{\rm I\kern-.18em R}}
\def\IT{\relax{\rm I\kern-.18em T}}
\def\ZZ{\relax{\sf Z\kern-.4em Z}}

\def\e{\epsilon} \def\G{\Gamma}   \def\k{\kappa}  \def\l{\lambda}
\def\L{\Lambda}     \def\si{\sigma}
\def\Si{\Sigma}

\def\cA{{\cal A}}  
\def\cD{{\cal D}}  
 \def\cH{{\cal H}} 
  
\def\cM{{\cal M}} \def\cN{{\cal N}} \def\cO{{\cal O}}

   \def\tF{{\tilde F}}

   \def\tphi{\tilde \phi}

\def\tPhi{\tilde \Phi}

 \def\wtN{{\widetilde N}}

 \def\wtPhi{{\widetilde \Phi}}

\def\rmBH{{\rm BH}}      
      \def\rmCHL{{\rm CHL}}    

\def\rmEM{{\rm EM}}     
    
    \def\rmGB{{\rm GB}}  \def\rmGL{{\rm GL}}     
    
\def\rmHet{{\rm Het}}     

           \def\rmIIA{{\rm IIA}}      \def\rmIm{{\rm Im}}

\def\rmML{{\rm ML}}      
   \def\rmMS{{\rm MS}}

\def\rmSL{{\rm SL}}      \def\rmSO{{\rm SO}}      \def\rmSp{{\rm Sp}}
    \def\rmSym{{\rm Sym}}

         \def\rmdet{{\rm det}}
\def\rmdiag{{\rm diag}}            
\def\rmdom{{\rm dom}}

\def\rmgcd{{\rm gcd}}         \def\rmgGB{{\rm gGB}}

\def\rmmic{{\rm mic}}

        \def\rmmod{{\rm mod}}

 \def\mathC{{\mathbb C}} 
  \def\mathN{{\mathbb N}}
 \def\mathQ{{\mathbb Q}} \def\mathR{{\mathbb R}}
\def\mathZ{{\mathbb Z}}

\def\fnote#1#2{\begingroup\def\thefootnote{#1}\footnote{#2}\addtocounter{footnote}{-1}\endgroup}

\def\beq{\begin{equation}}
\def\eeq{\end{equation}}
\def\bea{\begin{eqnarray}}
\def\eea{\end{eqnarray}}

\def\lleq#1{\label{#1}\eeq}
\let\nn=\nonumber
\def\tabroom{\hbox to0pt{\phantom{\Huge A}\hss}}
\def\notin{\ \hbox{{$\in$}\kern-.51em\hbox{/}}}

\def\ra{{\rightarrow}}
\def\lra{\longrightarrow}
\def\lolra{{\longleftrightarrow}}

 \def\vphi{\varphi}

  \def\E1Fq{E_1/\IF_q}

\def\notdiv{{\relax{~|\kern-.34em /~}}}

 \def\otau{{\overline{\tau}}}

\def\boxit#1{
\vbox{\hrule height1pt\hbox{\vrule width1pt\kern0.3cm
\vbox{\kern0.3cm\hbox{$\displaystyle#1$}\kern0.3cm}\kern0.3cm\vrule
width1pt}\hrule height1pt}}

\begin{document}

\hfill %{\bf CERN$-$PH$-$TH/2012$-$104}

%\hfill \today

\baselineskip=18pt
\parindent=0pt

\vskip .9truein

\centerline{\Large {\bf Automorphic Black Hole Entropy }}
 \vskip .1truein

\vskip .3truein

\centerline{{\sc~ Rolf Schimmrigk$^{\diamond}$}\fnote{}{$\diamond$netahu@yahoo.com, rschimmr@iusb.edu}}

\vskip .2truein

\centerline{Indiana University South Bend}
\centerline{1700 Mishawaka Ave., South Bend, IN 46634}

\vskip 1truein

\centerline{\bf Abstract}
\begin{quote}
 Over the past few years the understanding of the microscopic theory of black hole
 entropy has made important conceptual progress by recognizing that the degeneracies are
 encoded in partition functions which are determined by higher rank automorphic representations,
  in particular in the context of Siegel modular forms of genus two.
 In this brief review some of the elements of this framework are highlighted. One of the surprising
    aspects is that the Siegel forms that have appeared in the entropic framework are geometric in origin, arising
  from weight two cusp forms, hence from elliptic curves.
\end{quote}

\renewcommand\thepage{}
\newpage
\parindent=0pt

 \pagenumbering{arabic}

\baselineskip=15pt
\parskip=.02truein

\tableofcontents

\vskip .5truein

\baselineskip=21.5pt
\parskip=.15truein

\section{Introduction}\label{ra_sec1}

 Automorphic black hole entropy arises in the context of $\cN=4$ supersymmetric black hole entropy
 in theories with higher curvature couplings. Such black holes turn out to be a useful testing
 ground, in particular in the extremal limit, because they are simple enough, but not too simple.
 A typical case leads to actions of the schematic form
 \beq
  \cA ~=~ \int d^4x \sqrt{-g} \left( \phi_R(a_i) R ~-~ \phi_{IJ}(a_i) F^IF^J - \tphi_{IJ}(a_i)F^I\tF^J
       ~+~ \phi_\rmGB(a_i) \rmgGB  \right) + \cdots
 \lleq{action}
                                              %keep the minus sign in the $F^2$ term!
 where $\rmgGB$ collects fourth derivative curvature terms, e.g. the combination
 \beq
   \rmgGB ~=~ aR_{\mu \nu \k \rho}R^{\mu \nu \k \rho} + bR_{\mu \nu}R^{\mu \nu} + c R^2,
 \eeq
 which for $(a,b,c) = (1,-4,1)$ reduces to the Gauss-Bonnet term.
  The functions $\phi_\star$ describe the couplings of the scalar fields $a_i$
   of the theory and the dots indicate further terms not emphasized in the following discussion.
For actions of this type it is natural to expect that the black hole entropy decomposes
into two parts
 \beq
  S_\rmBH(Q) ~=~ S_\rmEM(Q) ~+~ S_\rmGB(Q)
 \eeq
 where $Q$ collectively denotes the charges associated to the gauge fields $F^I$, where $I$ is a
multiplet index.
 The first term gives the entropy due to the Einstein-Maxwell term,
 while the second term gives the dependence of the entropy on the higher curvature terms.

It turns out that these two parts of the action are quite different in structure: while the two-derivative
 action is quite insensitive to the details of the $\cN=4$ theory space, the  four-derivative
 higher curvature part not only depends on the detailed structure of the different models, it does so
 in a highly restricted way. The interesting phenomenon that occurs  in
 at least some classes of models
 is that the higher curvature coupling function $\phi_\rmGB(a_i)$,
  and therefore the resulting black hole entropy,
  is determined by functions $f$ that are modular with respect to Hecke congruence subgroups
  of the group $\rmSL(2,\mathZ)$, where the level of these forms is a characteristic of the
  theory. Modular forms are good because they are functions on the complex plane
  whose structure is extremely restricted by their symmetries. They are in fact
 so constrained that a finite
  number of computations determines them completely. Even more interesting is that the
 classical forms $f(\tau)$ that appear in certain models lift to automorphic forms $\Phi_f(\tau_i)$
 of higher rank with respect to the subgroups of the symplectic group and these automorphic forms
  determine the microscopic entropy. This framework therefore provides a tantalizing link
 between classical modular forms that appear in the effective action and higher rank automorphic
 forms that determine the microscopic partition functions for the entropy.

The appearance of such automorphic forms and their associated automorphic representations
 allows to link the framework of black hole entropy to the arithmetic Langlands program, in particular
 the reciprocity conjecture. The latter takes the point of view that automorphic forms and their
representations can be viewed as structures that are determined by representations of Galois groups,
 objects that are of a much simpler nature than automorphic representations.
 This is of interest because of Grothendieck's earlier
insight that Galois representations admit a geometric
interpretation in the framework of motives, geometric structures that can be viewed as the basic
 geometric building blocks of manifolds, defining "geometric atoms" in the original sense of the word.
Combining the three frameworks of automorphic black hole entropy, the arithmetic
 Langlands program, and  Grothendieck's theory of motives makes it natural to ask whether
  the modular forms that appear in  $\cN=4$ black hole entropy
admit a geometric interpretation. This will be discussed below.

The main goal of the present review is to provide an overview of the results obtained over
 the past years on the automorphic structure of black hole entropy in a certain class
 of $\cN=4$ models, with the aim to provide sufficient background material to illustrate the geometric
 origin of the resulting automorphic forms.
In order to keep the following discussion
within the confines of a brief review several interesting and important topics had to omitted, e.g.
 the problem of the moduli dependence of the entropy, in particular the phenomenon of
 wall crossing. Also omitted is  the class of black holes  with torsion.
 For more extensive discussions of not only these topics, but also
  concerning the specific details of the computation of $\cN=4$ entropies outlined here,
 the reader may consult the extensive review of Sen \cite{as07c},
  or the more recent shorter review by Mandal and Sen \cite{ms10}.
An in-depth analysis of the simplest model in this class
 can be found in the review of Dabholkar and Nampuri \cite{dn12}, while the useful summary
 of Mohaupt \cite{tm08} presents a slightly different point of view.

\section{Macroscopic $\cN=4$ black hole entropy}

Concrete results for $\cN=4$ black hole entropy have been known for some time, starting with
 the work of Dijkgraaf-Verlinde-Verlinde \cite{dvv96}
         (see also \cite{cdwkm04, ssy05a, g05, sy05, dmz12}),
 but more systematic results were obtained only  about a decade later by Jatkar-Sen \cite{js05}
 and Govindajaran-Krishna \cite{gk09} in the context
 of a special class of compactifications considered by  Chaudhuri-Hockney-Lykken \cite{chl95}.
  In the heterotic picture the models are constructed by considering
  toroidal quotients with respect to abelian discrete groups
 $\mathZ_N:=\mathZ/N\mathZ$ of order $N$, while in the dual IIA picture the compact spaces
 are quotients of products $K3\times T^2$ of K3 surfaces and elliptic curves $T^2$. The CHL$_N$
 models can therefore be viewed in the two dual pictures as
 \beq
  {\rm CHL}_N:~~~~\rmHet(T^6/\mathZ_N) ~\cong ~ \rmIIA((K3\times T^2)/\mathZ_N).
 \eeq
     (More details on CHL$_N$  dyons can be found in refs.
 \cite{cdwkm06, djs06a, dn06, ds06, djs06b, dg06, as07a,dgn07,as07b,cv07, dgm08, cd08, dgmn09, mp09, c10}.)
 In the context of $T^6$ quotients the action of $\mathZ_N$ factors
 into a transformation of order $N$ on the conformal field theory compactified on the 4-torus
 $T^4\subset T^6$ and an order $N$ shift along one of the remaining 1-cycles $S^1\subset T^6$.

The charge lattice $\L^{(N)}$
 depends on the order $N$ of the quotient group, leading to rank $r_N$ for the lattice associated
 to $\rmCHL_N$. The values of these $r_N$ are listed in Table 1.
 \begin{center}
 \begin{tabular}{l| c c c c c c c c}

 $N$   &1   &2  &3  &4   &5  &6  &7  &8 \tabroom \\
 \hline

 $r_N$ &28  &20 &16 &14  &12 &12 &10  &10  \tabroom \\

 \hline
 \end{tabular}
 \end{center}
 \centerline{{\bf Table 1.}~{\it Ranks $r_N$ of the CHL$_N$
 models.}}

 The electric and magnetic charges  $(Q_e,Q_m) = (Q_e^I, Q_m^I)$, $I=1,...,r_N$ arise from the
 vector multiplets associated to which are the moduli spaces given by the
 quotients
 \beq
 \cM(r_N) ~=~
 \rmSO(6,r_N-6,\mathZ) \backslash \rmSO(6,r_N-6,\mathR)/\rmSO(6,\mathR)\times \rmSO(r_N-6,\mathR)
 \eeq
 with respect to the discrete T-duality group $T_N=\rmSO(6,r_N-6,\mathZ)$ as well as the maximal compact
 subgroup. The T-duality invariant
 norms $Q_e^2$ for the electric charge, $Q_m^2$ for the magnetic charge, and $Q_eQ_m$ for their combination
 in turn can be combined to an S-duality invariant combination takes the form
 $Q_e^2Q_m^2-(Q_eQ_m)^2$. The S-duality group in $\cN=4$ supersymmetric theories is
in general a subgroup of the full modular group $\rmSL(2,\mathZ)$, a fact that was first
 emphasized by Vafa and Witten \cite{vw95} in the context of the CHL$_2$ quotient model.
 For general $N>1$ the relevant group has been identified in \cite{as05} to be
  given by the congruence subgroup
 $\G_1(N)$ of the modular group $\rmSL(2,\mathZ)$, defined as
 \beq
 \G_1(N)  = \left\{g\in \rmSL(2,\mathZ)~{\Big |}~ g ~\equiv ~
      \left(\matrix{1 &* \cr 0 &1 \cr}\right)(\rmmod~N)\right\},
 \eeq
  a subgroup of the Hecke congruence group $\G_0(N)$. A detailed discussion can be found in \cite{d07}.

Duality invariance thus makes it natural to expect that the black hole entropy
 in the lowest order of the effective theory takes the form \cite{cy95, ct95}
                %these are the references given by Dijkgraaf-Verlinde-Verlinde
 \beq
   S_\rmEM ~\cong ~ \pi \sqrt{Q_e^2Q_m^2 - (Q_eQ_m)^2}.
 \eeq
 Thinking of black holes as probes of the underlying theory, this leading order entropy is not very
 sensitive to the specific structure of the compactification variety $X_N$.
This lack of sensitivity of the Einstein-Maxwell part of the action makes it natural to consider
the effects of the higher curvature corrections in the action (\ref{action}).
 In the $\cN=4$ context corrections have been computed in refs.
  \cite{hm96, gkkopp97} and the first nonleading correction of the
curvature part of the action that has been considered is of the
 form (\ref{action}), with the standard Gauss-Bonnet term
 \beq
  \phi^{(N)} (a,S) \rmGB ~=~ \phi^{(N)}(a,S)
 \left( R_{\mu \nu \k \rho} R^{\mu \nu \k \rho} - 4R_{\mu \nu}R^{\mu \nu} + R^2\right),
 \eeq
 where  the Gauss-Bonnet coupling $\phi^{(N)}(S,a)$ is a function of the axion $a$ and the
 dilaton $S=e^{-2\vphi}$.  It is useful to write this coupling as a complex function
 $\phi^{(N)}(a,S) = \phi^{(N)}(\tau,\otau)$ in terms of the variable $\tau:=a+iS \in \cH$ in
 the upper half plane.

  The most interesting part of higher derivative action is that these couplings
are essentially determined by modular forms, i.e. functions $f$ on the upper half-plane
   $\cH \subset \mathC$ which under subgroups $\G \subset \rmSL(2,\mathZ)$ transform as
 \beq
  f(g\tau) ~=~ \e(d) (c\tau+d)^w f(\tau), ~~~g=\left(\matrix{a &b\cr c &d\cr}\right) \in \G,
 \eeq
 where $\e$ is a character and  the integer $w$ denotes the weight.
For each of the CHL$_N$ models the
 coupling $\phi^{(N)}$ becomes a function of the modular form
 $f^{(N)}(\tau)$. The structure of these forms is obtained by a simple rationale because
  covariance under the duality group is quite restrictive. This can be seen as follows \cite{cs11}.
 Recall first that the duality
 group $\rmSL(2,\mathZ)$ for the $N=1$ model is broken to a level $N$ group for
 $N>1$. For $N=1$ the result of Dijkgraaf-Verlinde-Verlinde \cite{dvv96} shows that
 the modular form is the discriminant form $f^{(1)}(\tau) = \eta(\tau)^{24}$, where
  $\eta(\tau) ~=~ q\prod_{n=1}^\infty (1-q^n)$ is the Dedekind eta function, written in terms
 $q=e^{2\pi i \tau}$. The function $f^{(1)}(\tau)$ is the unique modular cusp form
 of weight twelve and level one. Assuming that the forms $f^{(N)}$  for $N>1$ have level $N$
 leads for prime orders $N=p$ to unique candidate cusp forms that admit closed expressions as
 eta-products of the following type
 \beq
  f^{(N)}(q) ~=~ \eta(q)^{w+2} \eta(q^N)^{w+2} ~\in ~ S_{w+2}(\G_0(N),\e_N),
 \eeq
 where for $N=p=1,2,3,5,7$ the weight is determined as $w+2 = \frac{24}{(N+1)}$,
 and the modular groups are the Hecke congruence
 subgroups $\G_0(N) \subset \rmSL(2,\mathZ)$,  defined as
 \beq
 \G_0(N) ~:=~ \left\{g = \left(\matrix{a &b\cr c&d\cr}\right) \in \rmSL(2,\mathZ) ~{\Big |}~
   c\equiv 0(\rmmod~N)\right\}.
 \eeq
 Here the character $\e_N$ is trivial, except for $N=7$, for which it is given by the Legendre
 character $\e_7(d) = \chi_{-7}(d) = \left(\frac{-7}{d}\right)$. Legendre characters are defined as
 \beq
 \chi_N(p) ~=~ \left(\frac{N}{p}\right)
   ~=~ \left\{ \begin{tabular}{r l}
      $1$    &if  $x^2\equiv N(\rmmod~ p)$ is solvable \\
      $-1$  &if  $x^2\equiv N(\rmmod~ p)$ is not solvable \tabroom \\
       0 &if $p|N$. \tabroom \\
     \end{tabular}
  \right\}.
 \eeq

For the remaining composite values $N=4,6,8$ of the CHL$_N$ class of models the quotient
 $24/(N+1)$ is neither integral nor half-integral, hence there are no modular forms with corresponding
weights. It is natural to extend the prime sequence above by considering forms of weights
 \beq
  w_N ~=~ \left\lceil \frac{24}{N+1} \right\rceil
 \lleq{ceiling-weight}
 where $\lceil a \rceil$ denotes the next largest integral number obtained from the number $a$.
For $N=4,6,8$ these values lead to weights $5,4,3$, respectively. Assuming furthermore that the
 order $N$ of the quotient group $\mathZ_N$ again determines the level of the modular group
 leads to unique candidates forms given by eta-products. These forms and their
 characters, again given by Legendre symbols, are collected in Table 2. The uniquely determined
 forms obtain with the simple assumptions above are precisely the forms proposed by Jatkar and
 Sen \cite{js05} for prime orders and by Govindarajan and Krishna \cite{gk09} for composite orders.

%\begin{center}
\begin{tabular}{l | c c c c}

$N$        &$p\leq 7$ prime &4  &6  &8 \\
 \hline
 $f^{(N)}$
                   &$\eta(\tau)^{\frac{24}{p+1}} \eta(p\tau)^{\frac{24}{p+1}}$
                                 &$\eta(\tau)^4\eta(2\tau)^2\eta(4\tau)^2$
                                        &$(\eta(\tau)\eta(2\tau)\eta(3\tau)\eta(6\tau))^2$
                                                        &$\eta(\tau)^2 \eta(2\tau)\eta(4\tau)\eta(8\tau)^2$\tabroom \\
 \hline

 Type          &$S_w(\G_0(N),\e_N)$
                              &$S_5(\G_0(4), \chi_{-1})$
                                 &$S_4(\G_0(6))$
                                       &$S_3(\G_0(8), \chi_{-2})$ \tabroom \\

\hline

 \end{tabular}
%\end{center}

\vskip .2truein

\centerline{{\bf Table 2.}~{\it The classical modular forms of the $\rmCHL_N$ models.}}

\vskip .1truein

  The coupling $\phi^{(N)}(\tau,\otau)$ of the Gauss-Bonnet term
 then is essentially given by modular cusp forms $f^{(N)}(\tau)$ via
 \beq
   \phi^{(N)}(\tau, \otau)
 ~\cong ~  \ln ~\left[f^{(N)}(\tau) \cdot f^{(N)}(-\otau)\cdot (\rmIm~\tau)^{w+2}\right],
 \eeq
where $\rmIm~\tau$ is the holomorphic anomaly. For $N=1$ the resulting weight twelve form
 $f^{(1)}(\tau) = \Delta(\tau):= \eta^{24}(\tau)$ with respect to the full modular group $\rmSL(2,\mathZ)$
is familiar from the early days in string theory because
 it leads to the partition function of the bosonic string.

The next-to-leading order correction of the entropy beyond the Einstein-Maxwell system
 entropy  $S_\rmEM$ is determined in terms of charge expressions that do not change under the scaling
 \beq
  (Q_e,Q_m) ~\lra ~ (\l Q_e, \l Q_m)
 \eeq
 and are given by
 \bea
   Q^{(1)} &=& \frac{Q_eQ_m}{Q_m^2} \nn \\
   Q^{(2)} &=& \frac{1}{Q_m^2} \sqrt{Q_e^2Q_m^2 - (Q_eQ_m)^2}.
 \eea
 With $Q^{(i)}$ the corrected entropy term is of the form $S_\rmGB ~\cong ~ \phi^{(N)}( Q^{(1)}, Q^{(2)})$,
 i.e. the axion and dilaton pair $(a,S)$ are replaced by the corresponding charge expressions.
 More precisely, up to fourth order in the derivative expansion the entropy is given by
 \beq
   S_\rmBH(Q_e,Q_m) ~=~ S_\rmEM(Q_e,Q_m) + \phi^{(N)} \left(\frac{Q_eQ_m}{Q_m^2},~
      \frac{1}{Q_m^2} \sqrt{Q_e^2Q_m^2 - (Q_eQ_m)^2}\right),
 \eeq
 first implicitly derived implicitly in \cite{cdwkm04}, but most easily obtained
  via the entropy function formalism developed by Sen in several papers
  (e.g. in refs. \cite{as05b,as05c}),  and reviewed in detail in \cite{as07c}.

 This result for the Gauss-Bonnet contribution to the entropy shows that higher derivative corrections
 are much more sensitive to the details of the theory than the entropy based on the two-derivative
 action. This leads to the idea that if black holes are viewed as experimental objects they could
in principle be used as tools to test predictions of gravitational theories beyond Einstein's general
 relativity \cite{cs11}.  Entropy corrections arising from terms higher than fourth derivative order
 have been considered
 in ref. \cite{ah09}.

The computation of the entropy in the effective theory raises the question whether the
 contributions $S_E$ and $S_\rmGB$ admit a microscopic interpretation, i.e. whether
 there are functions that define microscopic degeneracies $d_\rmmic(Q)$ such that
 \beq
 S_\rmmic ~=~ \ln ~d_\rmmic
 \eeq
 produces the expressions above derived from the effective theory.

\section{General structure of automorphic black hole entropy}

As mentioned above,  over the past fifteen years or so impressive progress has been made toward
the resolution of a problem that is forty years old $-$ the
microscopic understanding of the entropy of black holes. It has proven useful to focus on black holes with extended
supersymmetries because this leads to black holes that are simple,
but not too simple. It was shown in particular that for certain types
of black holes in ${\cal N}=4$ supersymmetric theories  their
entropy is encoded in the Fourier coefficients of
Siegel modular forms, automorphic objects which provide one of the
simplest generalizations of classical modular forms of one
variable with respect to congruence subgroups of the full modular
group ${\rm SL}(2,{\mathbb Z})$.

A general conceptual framework of automorphic entropy functions
has not been established yet. A formulation that generalizes
the existing examples can be outlined as follows \cite{rs12}. Suppose
we have a theory which contains scalar fields parametrized by a
homogeneous space $\prod_i (G_i/H_i)$, where the $G_i$ are Lie
groups. Associated to these scalar fields are electric and
magnetic charge vectors $Q=(Q_e,Q_m)$, taking values in a lattice
$\Lambda$ whose rank is determined by the groups $G_i$.

Assume now that the theory in question has a T-duality group
$\prod_i D_i({\mathbb Z})$, where  the $D_i({\mathbb Z}) \subset
G_i({\mathbb Z})$ denote subgroups of the Lie groups $G_i$,
 considered over the rational integers ${\mathbb Z}$. Suppose further that the charge
vector $Q$ leads to norms $||Q||_i, ~i=1,...,r$ that are invariant
under the T-duality group. Choose conjugate to these invariant
charge norms complex chemical potentials
 \begin{equation}
 (\tau_i, ||Q||_i),~~ i=1,...,r,
 \end{equation}
which generalize the upper half plane of the bosonic string. On the generalized upper
 half plane ${\cal H}^r$ formed by the
variables $\tau_i$ one can consider automorphic forms
$\Phi(\tau_i)$, and the idea is that with an appropriate integral
structure ${\mathbb Z} \ni k_i \sim ||Q||_i, i=1,...,r$
 associated to the charge norms, the Fourier expansion of these
 automorphic forms given by
 \begin{equation}
  \Phi(\tau_i) ~=~ \sum_{k_n \in {\mathbb Z}} g(k_1,...,k_r)
                    q_{1}^{k_{1}} \cdots q_{r}^{k_{r}} ,
 \end{equation}
 in terms of $q_k = e^{2\pi i \tau_k}$,
 determines the automorphic entropy via the coefficients of the
 expansion of the automorphic partition function
 \begin{equation}
  Z(\tau_i) ~=~ \frac{1}{{\widetilde \Phi}(\tau_i)} ~=~ \sum_{k_n}
  d(k_1,...,k_r) q_{1}^{k_{1}} \cdots q_{r}^{k_{r}},
 \end{equation}
 as
 \begin{equation}
  S_{\rm mic}(Q) ~\cong ~ \ln d_\rmmic(Q).
 \end{equation}
  Here
 \begin{equation}
 d_\rmmic(Q) ~:=~ d(||Q||_1,...,||Q||_r)
 \end{equation}
 and ${\widetilde \Phi}$ denotes a modification of the Siegel form
 $\Phi$ that is determined by the divisor structure of $\Phi$.
 The precise definition of $\wtPhi$ is motivated by the interplay between the
 automorphic discrete group and the goal to isolate the dominant poles in
 $Z$.
If in leading order of the large charge expansion the degeneracies lead to the asymptotic result
 \beq
 d_\rmmic \cong ~e^{\pi \sqrt{F(||Q||_i)}},
 \eeq
 where $F(||Q||_i)$ is a quadratic form in terms of the norms $||Q||_i$, the microscopic entropy
 \beq
  S_\rmmic ~\cong ~ \pi \sqrt{ F(||Q||_i)}
 \eeq
 is structurally of the same type as the large charge limit of the macroscopic entropy.

\section{Siegel modular black holes in ${\cal N}=4$ theories}

The automorphic-entropy-outline of the previous section accounts for the behavior of
the entropy of black holes in certain ${\cal N}=4$
compactifications obtained by considering ${\mathbb
Z}_N-$quotients of the heterotic toroidal compactification
 ${\rm Het}(T^6)$, a small class of models first considered by
Chaudhuri-Hockney-Lykken models
 \cite{chl95}. Specifically, it was shown in \cite{dvv96,js05,gk09}
 that for these CHL$_N$ models the microscopic entropy of extreme
 Reissner-Nordstrom type black holes is
described by Siegel modular forms $\Phi^N \in
S_w(\Gamma^{(2)}_0(N))$, where the weight $w$ of $\Phi^N$ is determined by the
order $N$ of the quotient group. In this case the dyonic charges
$Q=(Q_e,Q_m)$ form three integral norms $||Q||_i$ invariant under the
T-duality group ${\rm SO}(6,r_N-6)$.
Associating to the norms $||Q||_i$  conjugate complex variables
 \beq
 (\tau,\si,\rho) ~=~ (\tau_1,\tau_2,\tau_3) ~~\lolra ~~
   (||Q||_1,||Q||_2,||Q||_3)
 \eeq
 leads to a three-dimensional domain for automorphic forms associated to
 CHL$_N$ models. This domain should generalize to dyonic black holes
 the upper half plane $\cH$
 on which the partition functions of purely electric and purely magnetic black holes
 are defined. For dimensional reasons $\rmGL(n)$-type automorphic forms are excluded,
 but Siegel-type automorphic forms of genus two are natural candidates because
 the Siegel upper half plane ${\cal H}_2$
 \begin{equation}
 T ~=~ \left(\matrix{
              \tau_1  &\tau_3 \cr \tau_3 &\tau_2\cr}\right)  \in {\cal H}_2,
 \end{equation}
with $\mathC \ni \tau, \si >0$ and $\rmdet ~\rmIm(T)>0$,  reduces in the degeneration
  $\tau_3 ~\ra~ 0$ into the  product of a
 pair of classical upper half planes
 \beq
  \cH_2 ~\stackrel{\rho \ra 0}{\lra}~ \cH_1\times \cH_1,
 \lleq{siegel-plane-degeneration}
  a fact
that will be reflected in the behavior of the automorphic forms.
 A key motivating result in this direction, pointing toward the usefulness of Siegel forms,
 is that in the simplest example, given by the Igusa form $\Phi_{10}$ of weight ten for the full
 symplectic group $\rmSp(4,\mathZ)$, the degeneration (\ref{siegel-plane-degeneration})
 leads to a factorization of the Igusa form as
 \beq
  \Phi_{10}(\tau,\si, \rho) ~~\stackrel{\rho \ra 0}{\lra} ~(2\pi i)^2 \rho^2 \Delta(\tau) \Delta(\si),
 \eeq
 where $\Delta(\tau)$ is precisely the modular form of weight twelve that appears in the
 Gauss-Bonnet coupling $\phi^{(N)}(\tau,\otau)$ of the effective action for the CHL$_1$ model.

 The automorphic groups relevant for the CHL$_N$ models
 are Hecke type genus two congruence subgroups
 $\Gamma_0^{(2)}(N)$ of the symplectic group
 \beq
 {\rm Sp}(4,{\mathbb Z})
  ~=~ \left\{M = \left(\matrix{A &B\cr C &D\cr}\right) ~{\Big |}~ M^tJM = J,
      ~J = \left(\matrix{~~0 &{\bf 1}\cr - {\bf 1} &0 \cr}\right) \right\},
 \eeq
 hence the associated
 forms are Siegel modular forms of genus two,  i.e. functions $\Phi$ on $\cH_2$ which
transform with respect to $\G_0^{(2)}(N) \subset \rmSp(4,\mathZ)$
in a way analogous to classical modular forms.
 A genus two Siegel modular form $\Phi_w$ is said to be of weight $w$ if for any
 \beq
 M ~=~ \left(\matrix{A &B\cr C &D\cr}\right)~ \in~ \G_0^{(2)}(N)
 \eeq
 with $(2\times 2)$-matrices $A,B,C,D$ it transforms under
 \beq
   T ~\longmapsto ~ MT := (AT+B)(CT+D)^{-1}
 \eeq
as
 \beq
 \Phi_w(MT) ~=~ \rmdet(CT+ D)^w \Phi_w(T).
 \lleq{siegel-transform}

For the CHL$_N$ models the weight $w$ of the associated Siegel form $\Phi^{(N)}$ is given by
 \beq
  w(\Phi^{(N)}) ~=~ w_N-2,
 \eeq
 where $w_N$ is the weight defined in (\ref{ceiling-weight}).

 The Siegel forms $\Phi^{(N)}$ do not immediately define the partition functions in the models
 for $N>1$. The problem that arises is as follows. First one notes that already for the
 $N=1$ CHL model the diagonal divisor
\beq
  \cD_\rmdiag=\{\rho^2=0\},
  \lleq{diagonal-divisor}
 which arises in the limit $\rho ~\ra~ 0$ mentioned above via the factorization of the Siegel form
      \beq
     \Phi^{(N)}(\tau,\si,\rho) ~~\stackrel{\rho \ra 0}{\lra} ~\sim~ \rho^2 f^{(N)}(\tau)f^{(N)}(\si)
    \eeq
 does not provide the dominant contribution to the entropy, but is suppressed in the large charge
 expansion of the degeneracies $d(k,\ell,m)$.
The leading  contribution  instead is given by the dominant divisor
  \beq
   \cD_\rmdom = \{\rho^2-\rho -\tau \si =0\},
 \lleq{dominant-divisor}
  which can be obtained as an $\rmSp(4,\mathZ)$ image of $\cD_\rmdiag$.
 In the $N=1$ model the symplectic group $\rmSp(4,\mathZ)$
  is the symmetry group of the theory, hence the Siegel form
 can be transformed as (\ref{siegel-transform}) for such a group element. For the models $N>1$
 the map considered in \cite{js05} is however not an element of the symmetry group $\G_0^{(2)}(N)$,
 hence the Siegel forms $\Phi^{(N)}(T)$ do not transform under this map. The practical way out  is
 to introduce functions $\tPhi^{(N)}(T)$ in precisely such a way that the large charge limit
 agrees with the macroscopic theory. This leads to the definition of the Siegel type partition function
 \beq
  Z(\tau_i) ~=~ \frac{1}{\tPhi(\tau_i)},
 \lleq{siegel-partition-function}
 with  $\tPhi(\tau_i)$, obtained from $\Phi(\tau_i)$ by multiplication
 with a function $A(\tau,\si,\rho)$ determined by the map from the diagonal to the dominant divisor.
 The motivation for this becomes clear already in the analysis of the $N=1$ case.

In the $N=1$ model $\rmHet(T^6)$ the Siegel form must transform with respect to the
  full symplectic group $\rmSp(4,\mathZ)$ of genus two with weight $w=10$.
 This uniquely determines the form $\Phi_{10}$
 to be the Igusa modular form, which leads directly to the degeneracies that enter the entropy by expanding
 \beq
  Z(\tau, \si,\rho) ~=~ \frac{1}{\Phi_{10}(\tau,\si, \rho)}
              ~=~ \sum_{k,\ell,m} d(k,\ell,m) q^k r^\ell s^m,
 \lleq{siegel-fourier}
 where $q=e^{2\pi i \tau}, r=e^{2\pi i \si}, s=e^{2\pi i \rho}$
  (for $N>1$ the divisor induction modification $\tPhi$ has to be considered). With  \cite{sy05}
   %for the sign!
 \beq
  d(Q) ~=~ (-1)^{Q_eQ_m+1} \oint dT ~\frac{e^{-\pi i Q^t T Q}}{\Phi_w(T)}
\lleq{degeneracy-integral}
 the degeneracies
 of the (in general modified) inverse Siegel form define the microscopic entropy
 \beq
   S_\rmmic(Q) ~=~ \ln ~d_\rmmic(Q).
 \eeq
 The integral (\ref{degeneracy-integral})  can be evaluated
by computing first the Cauchy integral
 for the off-diagonal variable for the pole given by the Humbert divisor and then evaluating the
 remaining integral in a saddle point approximation.

Insight into how contact with the
 macroscopic entropy is made can be obtained without going through the whole computing
 by simply making explicit the result of the $\rho$-integral obtained by using the transformation
  (\ref{siegel-transform}) for the matrix
  ${\tiny M=\left(\matrix{A &B\cr C &D\cr}\right)}$ mapping the diagonal
divisor (\ref{diagonal-divisor}) to the dominant divisor (\ref{dominant-divisor}),
given in terms of the block
 matrices as $A={\tiny \left(\matrix{1 &0 \cr 0 &0\cr}\right),  B=\left(\matrix{ 0 &0  \cr0 &1 \cr}\right),
   C=\left(\matrix{0 &~0 \cr 1 &-1\cr}\right), D=\left(\matrix{1&1\cr 0 &0\cr}\right)}$.
 With $T'=MT$ one gets for a weight $w$ Siegel form $\Phi_w(T') = \rmdet(CT+D)^w \Phi_w(T)
   ~=~ (\si')^{-w} \Phi_w(T)$. Taking the factorization limit in the $T$-variables and mapping with
 $M$ the diagonal divisor $\rho^2$ to the primed coordinates gives essentially the dominant divisor,
 since $\rho=-(\rho'^2-\rho'-\tau'\si')/\si'$. These maneuvers then lead to the transformed integral
 \beq
  d_\rmdom(Q) ~\cong ~ \int d\tau' d\si' d\rho'
     ~\frac{(\si')^{w+2}e^{-\pi Q^tT'Q}}{(\rho'^2-\rho' -\tau'\si')^2 \Delta(\tau)\Delta(\si)},
 \eeq
 where $(w+2)=12$ is the weight of the form $\Delta$
 and  $\tau =(\tau' \si'-\rho'^2)/\si'$ and $\si=(\tau'\si' - (\rho'-1)^2)/\si'$
arise from the transformation given by $M$. The residue integral leads to an integral
 of the form
\beq
  d_\rmmic(Q) ~\cong ~ \int d\tau' d\si' ~e^{-2\pi i \Si^\rmmic(\tau',\si')} J(\tau',\si'),
 \eeq
 where the microscopic entropy function $\Si^\rmmic(\tau,\si)$ and $J(\tau,\si)$
 are somewhat unwieldy expressions, but the key is that $\Si^\rmmic$
 essentially contains a term
 \beq
 \psi(\tau',\si') ~\cong~ \ln ~\Delta(\tau)\Delta(\si) ~(\si')^{-(w+2)},
 \eeq
  which looks reminiscent to the expression $\phi^{(1)}$ of the higher derivative effective action,
 and in fact becomes identical to $\phi^{(1)}$ in the saddle point evaluation.

In leading order of the large charge expansion the remaining saddle point evaluation leads to
 \beq
  d_\rmmic(Q) ~\cong ~ e^{\pi \sqrt{Q_e^2Q_m^2-(Q_eQ_m)^2}},
 \eeq
 in agreement with the macroscopic entropy described above  \cite{dvv96}.
 In subleading order the agreement between the macroscopic and microscopic entropy has been shown for the
 Gauss-Bonnet term in ref. \cite{cdwkm04} for the $N=1$ model. The generalization to $N>1$ has been discussed
  in \cite{js05} for prime orders, and for the remaining composite orders in \cite{gk09} and the issue
 of the path dependence of the degeneracy integral has been addressed in \cite{as07a, dgn07, as07b}.
The above outline for a microscopic interpretation of CHL$_N$ black hole entropy is valid for
charge configurations which satisfy the constraint
 \beq
  t_Q := \rmgcd\{Q_e^I Q_m^J - Q_e^J Q_m^I ~{\Big |}~ 1\leq I,J \leq r_N\} ~=~ 1.
 \eeq
 The integer $t_Q$ is called the torsion of the black hole, and  the
 issue of $\cN=4$ black holes with nontrivial torsion $t_Q>1$ was first raised in \cite{dgn07},
 and further discussed in several papers \cite{bs08,bss08a, bss08b, dgm08}.
The extensions and issues briefly mentioned above are reviewed in detail refs. \cite{as07c,ms10} where
 further references can be found.

\section{From black hole Siegel forms to higher weight classical modular forms}

The key feature of the Siegel modular forms that appear in the
context of CHL$_N$ black hole entropy is that they are not of
general type, but belong to the Maa\ss~Spezialschar, more precisely they are
obtained via a combination of the Skoruppa lift \cite{nps92}, which maps
classical modular forms to Jacobi forms, and the Maa\ss~lift \cite{m79}, which maps
 Jacobi forms to Siegel modular forms
 \begin{equation}
  f(\tau) \in S_{w+2} ~\stackrel{\rm SL}{\longrightarrow} ~\varphi_{w,1}(\tau,\rho) \in J_w
                      ~\stackrel{\rm ML}{\longrightarrow} ~\Phi_w(\tau,\sigma,\rho) \in S_w,
 \end{equation}
 where $\tau=\tau_1, \sigma=\tau_2, \rho=\tau_3$. Here the Maa\ss-lift ML sends a
 Jacobi form $\vphi_{w,1}$ of weight $w$ and level 1 with a Fourier expansion
 \beq
 \vphi_{w,1}(\tau,\si) ~=~ \sum_{\stackrel{k\in \mathN_0, \ell \in \mathZ}{4k-\ell^2\geq 0}}
     c(k,\ell) q^k r^\ell
 \eeq
 to a Siegel form of weight $w$ with the Fourier expansion
 \beq
  \Phi_w(q,r,s) ~=~  \sum_{\stackrel{k,\ell \in \mathN_0, m \in \mathZ}{4k\ell-m^2\geq 0}}
       g(k,\ell,m) q^kr^\ell s^m
\eeq
 with coefficients
 \beq
   g(k,\ell,m) ~=~ \sum_{d|(k,\ell,m)} \chi(d) d^{w-1} c\left(\frac{k\ell}{d^2},\frac{m}{d}\right),
 \eeq
 with Legendre character $\chi$.

The Skoruppa lift SL is a map that sends classical forms $f\in S_w(\G_0(N), \e_N)$
of weight $w$, level $N$, and character $\e_N$ to Jacobi forms
 via the prime form
 \beq
  K(\tau,\si) ~:=~ \frac{\vartheta_1(\tau,\si)}{\eta^3(\tau)}
 \eeq
 given in terms of the theta series
 \beq
  \vartheta_1(q,s) ~=~ \sum_{n\in \mathZ} (-1)^n q^{\frac{1}{8}(2n+1)^2} s^{n+\frac{1}{2}}
 \eeq
 and the Dedekind eta function defined above as
 \beq
  \rmSL(f) ~=~ \vphi_{w,1} := K^2f.
\eeq
 The combination
\beq
  \rmMS ~=~ \rmML \circ \rmSL
 \eeq
 defines a map from classical cusp forms $f(\tau)$ of weight $(w+2)$ to Siegel forms of genus two
 of weight $w$.
The form $f(\tau)$ whose Maa\ss-Skoruppa lift is the Siegel modular form
 $\Phi_w = {\rm MS}(f)$ is called the Maa\ss-Skoruppa root.

\section{Geometric origin of automorphic black hole entropy}

The fact that the automorphic black hole entropy of CHL$_N$ models encoded in the Siegel
 forms $\Phi^{(N)}$ is sensitive to the details of the theory, not only through the spectrum but also
 through the couplings, motivates the question what exactly the essential information is that is
 contained in these entropy functions. Put differently, the issue becomes what the irreducible structure
is that determines the entropy of these models. This question can be raised independently of any
specific picture in which the Siegel forms $\Phi^{(N)}$ are constructed, be that
  via type II D-branes, or M-theory branes,
 or with other ingredients provided by different dual pictures. The
point here is that any given construction can in principle introduce redundant structures and in the process
 might  not point to the irreducible building blocks. In the following this question
is motivated in the first subsection in a more general framework, while the second subsection constructs
 the irreducible building block of the CHL$_N$ models, following \cite{cs11}.
 The second subsection is structurally independent of the first.

\subsection{Historical background}

It has been known for more than a century that there is a connection between certain types
 of modular forms and certain types of  geometric structures. This insight can be traced to the work
 of Klein and Fricke, and in more recent times this observation has been generalized in both the
 geometric and and the automorphic direction. The first generalization makes the
 geometric aspects much more detailed by refining the monolithic geometries
 considered by Klein, Fricke, Hecke, Eichler and Shimura into the much more detailed and precise
 motivic framework of  Grothendieck \cite{g60s}. The second extension generalizes the  concept of
 classical modular forms, associated to subgroups of the modular group $\rmSL(2,\mathZ)$ acting
 on the upper half plane, to the notion of automorphic forms.  The latter can be viewed as
 functions on higher dimensional half planes, or as functions on linear algebraic
 reductive groups. The link between these a priori completely
 independent objects, i.e. motives defined by algebraic cycles on the one hand, and automorphic
forms and representations on the other, is provided by the concept of a Galois representation.
 It is in this context that the arithmetic Langlands program enters: the reciprocity
 conjecture posits a general relation between Galois representations and automorphic representations
 \cite{rl70}. Combining Langlands' reciprocity conjecture with Grothendieck's description of motives
as Galois representations makes it possible to think of motives as carrying an automorphic
structure. The direction from automorphic forms to motives is less clear because not all automorphic
forms are motivic. It is generally expected though that the subclass of algebraic automorphic
 forms are in fact motivic.

The idea of automorphic motives makes it natural to ask whether the automorphic forms that
 appear in the context of $\cN=4$ black hole entropy have a geometric interpretation. This question
 can be raised independently of the picture associated to the purported geometry: whether this is interpreted
 as a motive that lives in the compact direction of spacetime or whether it is interpreted as (part of) a
 moduli space. In the present section the focus is on establishing a geometric interpretation
 of the black hole entropy, while the final section puts this result into a broader perspective by
 summarizing the general motivic structure which is conjectured to be associated to Siegel modular forms.

\subsection{Lifts of weight two forms}

The key to the identification of the motivic origin of the CHL$_N$
black hole entropy turns out to be an additional lift construction
that interprets the Maa\ss-Skoruppa roots of weight $(w+2)$ in
terms of modular forms of weight two for all $N$, and hence in
terms of elliptic curves \cite{cs11}. These Maa\ss-Skoruppa roots
decompose into two distinct classes of forms, one class admitting
complex multiplication, the second class not. For this reason it
is natural to expect that the lifts of weight two modular forms to
the CHL$_N$ Maa\ss-Skoruppa roots involve two different
constructions, depending on the type of the higher weight form.
For forms without complex multiplication the lift
interpretation of the MS root $f_{w+2}$  in terms of the weight
two form $f_2 \in S_2$ can be written as
 \begin{equation}
  f_{w+2}(q) ~=~ f_2(q^{1/m})^m, ~~~~{\rm with}~~m =
  \frac{1}{2}\left\lceil \frac{24}{N+1}\right\rceil.
  \lleq{noncm-lift}
 The relation between the levels $\wtN$ of the weight two forms $f_2$ and the order $N$
  of the defining group $\mathZ_N$ is made explicit in Table 3.
 \begin{center}
 \begin{tabular}{l| c c c c c}
 Order $N$        &1 &2 &3 &5 &6 \tabroom \\
 \hline

 Level $\wtN$ &36 &32 &27  &20 &24 \tabroom \\

 \hline
 \end{tabular}
 \end{center}

 \centerline{{\bf Table 3.}~{\it The levels $\wtN$ in terms of
 the orders $N$ of the CHL$_N$ models.}}

For the CHL$_N$ models with $N=4,7,8$ the lift  (\ref{noncm-lift}) cannot be
applied because the MS roots $f_{w+2}$ have odd weight. It is therefore necessary
to come up with a different type of reduction. Inspection of the forms $f_{w+2}$
 shows that they admit complex multiplication. Intuitively, this means that the
Fourier expansion of these functions are sparse. The vanishing of the Fourier coefficients
 $a_p$ for an infinite number of primes $p$ is determined by the splitting behavior of
these primes in an associated number fields. The coefficients $a_p$ vanish for precisely
 those primes $p$ that do not split in the ring of integers $\cO_{K_D}$, where $K_D=\mathQ(\sqrt{-D})$
 is an imaginary quadratic field. This splitting behavior is controlled by the Legendre symbol $\chi_D$
and therefore a complex multiplication form $f\in S_w(\G_0(N),\e_N)$ can be defined through its
Fourier expansion by the condition that there exists a field $K_D$ such that
 \beq
 \chi_D(p) a_p ~=~ a_p.
 \eeq

 It is useful to change the point of view and consider instead of the Fourier series $f(q) = \sum_n a_nq^n$
 of the form its associated $L$-series
 \beq
  L(f,s) ~=~ \sum_n \frac{a_n}{n^s},
 \eeq
 where $s$ is a complex variable.
 The lift for the class of MS roots with complex
 multiplication derives from the existence of algebraic Hecke
 characters $\Psi$ whose $L-$functions are the inverse Mellin transform
 of the MS roots
 \beq
  L(f,s) ~=~ L(\Psi,s).
 \eeq
 More details can be found in ref. \cite{cs11}.
The weight two forms that correspond to the classical higher
  weight forms $f_{w+2}$ are described in Table 4.

 \begin{center}
 \begin{small}
 \begin{tabular}{c |  c l | c l c}

 $N$ of     &   &BH Form       &   &Motivic form       &Level $\wtN$  \\
 CHL$_N$    &   &$f^N(q) \in S_{w+2}(\G_0(N))$
                               &   &$f_2^\wtN(q)$      &of $E_{\wtN}$ \\

 \hline
  1         &   &$\eta(\tau)^{24}$
                               &CM    & $\eta(q^6)^4 \in S_2(\G_0(36))$
                                                       &16  \tabroom  \\

 \hline

  2         &    &$\eta(\tau)^8\eta(2\tau)^8$
     &CM     &$\eta(q^4)^2 \eta(q^8)^2 \in S_2(\G_0(32))$  &32  \tabroom  \\
    %     &          &CM: $\eta(q^4)^2 \eta(q^8)^2\otimes \chi_2 \in S_2(\G_0(64))$ & 64 \\

 \hline

  3         &   &$\eta(\tau)^6\eta(3\tau)^6$
                       &CM   &$\eta(q^3)^2 \eta(q^9)^2$
                                     &27 \tabroom \\

  \hline

  4         &CM  & $\eta(\tau)^4\eta(2\tau)^2\eta(4\tau)^4$
                           &   &$\rmSym^4(f_2^{32})$ with $f_2^{32} \in S_2(\G_0(32))$
                            & 32  \tabroom \\

  \hline

  5         &   &$\eta(\tau)^4\eta(5\tau)^4$
                           &  &$\eta(q^2)^2 \eta(q^{10})^2 \in S_2(\G_0(20))$
                           &20  \tabroom   \\

  \hline

  6         &  &$(\eta(\tau)\eta(2\tau)\eta(3\tau)\eta(6\tau))^2$
       &  &$\eta(2\tau)\eta(4\tau)\eta(6\tau)\eta(12\tau) \in S_2(\G_0(24))$
                          & 24 \tabroom  \\

  \hline

  7         &CM  &$\eta(\tau)^3\eta(7\tau)^3$
         &    &$\rmSym^2f_2^{49}$ with $f_2^{49} \in S_2(\G_0(49))$
                                    & 49 \tabroom \\

 \hline

  8          &CM  &$\eta(\tau)^2\eta(2\tau)\eta(4\tau)\eta(8\tau)^2$
         &   &$\rmSym^2f_2^{256}$ with $f_2^{256} \in S_2(\G_0(256))$
                                    & 256 \tabroom \\

 \hline
 \end{tabular}
 \end{small}
 \end{center}

\centerline{{\bf Table 4.}~{\it Elliptic motives associated to the electric
modular forms of CHL$_N$ models.}}

 The interpretation of the Maa\ss-Skoruppa roots in terms of weight 2 modular
 forms $f_2^{\tilde N}$ via these two additional lifts for CM and non-CM forms
 shows that the motivic orgin of the Siegel modular
 entropy of CHL$_N$ models is to be found in elliptic curves. This follows
 from the fact that for all CHL$_N$ models the geometric structure that
 supports the weight 2 forms is that of elliptic curves $E_{\tilde N}$,
 whose conductor ${\tilde N}$ varies with the order $N$ of the quotient
 group ${\mathbb Z}_N$. More precisely, the $L-$functions associated to both of
 these objects agree
  \begin{equation}
   L(f_2^{\tilde N},s) ~=~ L(E_{\tilde N},s).
  \end{equation}
 Abstractly, this follows from the proof of the
 Shimura-Taniyama-Weil conjecture \cite{w95, tw95, bcdt01},
 but no such heavy machinery is necessary for the concrete cases
 based on the CHL$_N$ models, where the elliptic curves can be
 determined explicitly for each $N$.
  This shows that the motivic origin of the
 Siegel black hole entropy for the CHL$_N$ models can be reduced to that of
 complex curves $E_\wtN$. It is the arithmetic structure of these elliptic curves that carries
 the essential information of the entropy. (A detailed analysis of this arithmetic structure
 of elliptic curves in a physical context can be found in ref. \cite{rs05} and applications
 of elliptic curves as building blocks of Calabi-Yau threefolds appear in \cite{rs08}.)

\section{Automorphic motives}

 The general framework of automorphic motives raises the natural question whether it is
possible to provide a direct motivic interpretation of the Siegel modular forms that
 encode the microscopic nature of $\cN=4$ black hole entropy in the context of
 $\rmCHL_N$ models.  This would immediately lead to the picture of using black holes
 to extract geometric information  if we were able to
experiment with them in the laboratory.  If the resulting motives were of spacetime
 origin one might expect that such
automorphic black holes encode information about the geometry
of the extra dimensions in
string theory. This raises the question of how one can identify
the motives of the variety which support the automorphic forms
that appear in the entropy results.

Given that the entropy of black holes is described by automorphic
forms, one can ask whether the spacetime structure of the
compactification manifolds leads to motives which could support
these automorphic forms. It is not expected that general
automorphic forms are of motivic origin, however algebraic
automorphic forms are conjectured to be motivic.
 Background material for Siegel forms can be found in \cite{a09} and discussions of
  their conjectured motivic structure can be found in
 \cite{y01, p11}.
In the special case of the genus two Siegel modular forms
that appear in the context of CHL$_N$ black holes the conjectures
concerning the motivic origin indicate that the compactification
manifold cannot provide  the appropriate motivic cycle
structure in the way envisioned in the Siegel motivic literature.
The easiest way to see this is as follows \cite{cs11}.
Suppose $M_\Phi$ is a motive whose $L-$function $L(M_\Phi,s)$
agrees with the spinor $L-$function $L_{\rm sp}(\Phi,s)$ associated
to a Siegel modular form $\Phi$ of arbitrary genus $g$ and weight
$w$
 \begin{equation}
  L(M_\Phi,s) ~=~ L_{\rm sp}(\Phi,s).
 \end{equation}
 The weight ${\rm wt}(M_\Phi)$ of such genus $g$ spinor motives follows from the (conjectured)
 functional equation of the $L-$function as
 \begin{equation}
  {\rm wt}(M_\Phi) ~=~ gw - \frac{g}{2}(g+1).
 \end{equation}
 For the special case of genus 2 spinor motives the Hodge
 structure takes the form
 \begin{equation}
  H(M_\Phi) ~=~ H^{2w-3,0} \oplus H^{w-1,w-2} \oplus
                H^{w-2,w-1} \oplus H^{0,2w-3}.
 \label{siegel-hodge}\end{equation}
 This Hodge structure only applies to pure
 motives. In the case of mixed motives it is possible, for example,
 that rank 4 motives can give rise to classical modular forms  \cite{kls10}.

 While the Hodge type  (\ref{siegel-hodge})  of $M_\Phi$ is that of a Calabi-Yau variety,
 the precise structure is only correct for modular forms of weight
 three. Inspection shows that for the class of CHL$_N$ models the
 weights of the Siegel modular forms take values in a much wider
 range $w \in  [1,10]$.
 If follows that for most CHL$_N$ models the Siegel modular form
 will be of the wrong weight to be induced directly by motives in
 the way usually envisioned in the conjectures of arithmetic geometry.

The same is the case for the classical Maa\ss-Skoruppa roots,
whose weights are given by $(w+2)$. The motivic support $M_f$ for
such modular forms $f$ is of the form
 \begin{equation}
 H(M_f) ~=~ H^{w-1,0} \oplus H^{0,w-1},
 \end{equation}
 hence the only modular forms that can fit into heterotic
 compactifications have weight two, three, or four.
 This fact motivates the attempt to construct the Maa\ss-Skoruppa roots in terms
 of the simplest possible geometric modular forms, namely elliptic modular forms of weight two.
 This can be done as described in the previous section \cite{cs11}.

\vskip .1truein

 \noindent
{\bf Acknowledgement}

 Parts of this review are based on work that was conducted at the Max Planck Institute for Physics.
 It is a pleasure to thank Dieter L\"ust for making my visit possible, and the Werner Heisenberg
 Institut f\"ur Physik for hospitality and support.
 I'm also grateful to Wolfgang Lerche and CERN for supporting a visit during which I benefitted
 from discussions in particular with Atish Dabholkar and Stefan Hohenegger.
 It is a pleasure to thank Ilka Brunner,  Rajesh Gopakumar, Monika Lynker, Boris Poline,
 and Stephan Stieberger for further discussions.
 The work reviewed here was supported in part by the National Science Foundation
 under grant No. PHY 0969875.

%\vfill \eject

\end{document}